# The Impact of Computing Data Centres Orbiting Earth

Geoffrey W. Marcy[1*]

[1] *Space Laser Awareness, 3388 Petaluma Hill Rd, Santa Rosa, CA, 95404, USA*

**ABSTRACT**

Artificial intelligence is projected to increase U.S. data-centre power demand beyond 100 gigawatt (GW) by 2035 and global demand toward 1 terrawatt. In response, companies and governments have proposed placing computing infrastructure in sun-synchronous low-Earth orbit, where continuous sunlight could supply electrical power. Generating 5 GW would require solar arrays 4 × 4 km in size. Although technically feasible, such kilometre-scale structures at roughly 500 km altitude would dramatically alter both the night and daytime sky. A 4 × 4 km array in low-Earth orbit would span about 0.4°, comparable to the Moon, and reflected sunlight would make it shine at magnitude $g \approx -5$ to $-7$ mag, 100 times brighter than the brightest stars. Dozens of these structures would appear as a north–south chain of industrial objects across the sky, visible for about 1 1/2 hours after sunset and 1 1/2 hours before sunrise. They would block stars, planets, and deep-sky objects for minutes at a time, while increasing the likelihood of collisions that could trigger runaway debris production. These orbiting computing facilities therefore pose serious astronomical, technical, and cultural concerns.

## 1   INTRODUCTION

The explosive growth of artificial intelligence is driving a parallel surge in electricity demand. In the United States alone, data centres may require more than 100 gigawatts of power by 2035, while global AI-related demand could eventually approach 1 terawatt. In response, several companies and government-backed groups have proposed moving computing infrastructure into orbit, powered by large solar arrays.

The concept is technically attractive. In space, solar panels can receive nearly continuous sunlight, allowing far greater energy generation than ground-based systems. Spacecraft can radiate waste heat into space, reducing some of the cooling burdens that constrain terrestrial data centres, although important technical challenges remain. In addition, laser links between spacecraft could reduce dependence on much of the physical internet infrastructure required on the ground. From an engineering perspective, orbital AI computing offers a plausible path to large-scale power and high-capacity data connectivity.

Proposals to place computing infrastructure into low-Earth orbit (LEO) are proliferating. SpaceX has filed with the FCC for one million solar-powered "orbital data-center" satellites, specifying 100 GW of AI computing capacity per year (SpaceX 2026; *Data Center Dynamics* 2026). The startup Starcloud has applied to the FCC for a constellation of up to 88,000 satellites (Foust 2026a). Blue Origin has filed with the FCC to deploy 51,600 satellites for data-center operations (Foust 2026b). Google's Project Suncatcher plans constellations of AI-chip satellites, with launches proposed as early as 2027 (Beals et al. 2025a,b). China's Adaspace is pursuing similar space-based data-center concepts (Russell 2025; CASC 2026). Here we determine the size, brightness, and appearance of these computing centres in orbit.

## 2  GEOMETRY AND SIZE OF DATA-CENTER SATELLITES

Current proposals envision data centers orbiting at an altitude of ~500 km above the Earth's surface (LEO). They will reside in a Sun-synchronous orbit in which the orbital plane is perpendicular to the direction toward the Sun, thereby remaining bathed in solar flux all the time (Figure 1). As the Earth orbits the Sun, the satellites reorient their orbits, by ~1 deg per day, to remain constantly in sunlight.

The size of the solar panels required to generate power of 5 GW is easily calculated from the Sun's energy flux of 1,361 W m$^{-2}$, yielding a required area of 15 km$^2$, assuming a typical 25 percent efficiency. That area corresponds to a square patch in space 3.9 x 3.9 km, likely configured as rectangular arrays. Most current satellites in LEO, such as Starlink, have solar panels with an area of only 105 m$^2$. Thus the proposed 15 km$^2$ solar panels will have 140 000 greater cross-sectional area than today's satellites. They will reflect ~140 000 times more sunlight.

The kilometer-size solar panels and large bus structures that house the compute data-centers would form extended objects moving across the sky. In Sun-synchronous orbits, the data centers would populate a swath across the sky, extending from due north to due south. At both sunset and sunrise, the swath of kilometer-size satellites would pass directly overhead and be bathed in direct sunlight, crossing the sky.

Figure 1 shows the geometrical properties of the Sun-synchronous orbits of the data centers as seen by an observer after sunset. During the 1 ½ hrs after sunset, an observer at mid-latitudes is carried eastward by the Earth's rotation. The observer will see the data-centres appear to migrate from overhead toward the west, tracking the already-set Sun.

The time for the band of sun-synchronous satellites to migrate from overhead to the western horizon after sunset is easily shown from trigonometry in Figure 1 to be ~90 minutes after sunset. The Earth's radius is $R = 6371$ km, implying that the orbital radius is given by, $r = R + 500$ km $= 6871$ km. After the Sun has set, the sun-synchronous satellites (initially overhead) appear to move toward the western horizon. When they set on the observer's horizon (Fig. 1) the angle between line joining the centre of the Earth to the observer and the line from the centre to a satellite on the observer's horizon is $\alpha = \arccos(R/r) = 22$ deg, as shown in Figure 1. Since the Earth rotates eastward at 15 deg/hr, the 22 deg of Earth's rotation took ~90 min. During 90 min, data centres appear to move from overhead to the horizon – visible the entire time.

The exact time a sun-synchronous data center will set in the west after sunset depends on latitude and refraction in the Earth's atmosphere (Fig. 1). At higher latitudes the data centers will be visible longer. In short, at sunset the swath of kilometer-size data centers will appear as an arch passing overhead, from north-to-south. During the next 90 minutes, the arch of satellites will appear to descend westward, setting beneath the western horizon. The data centers will also be visible during the 90 minutes prior to sunset and for 90 minutes both before and after sunrise.

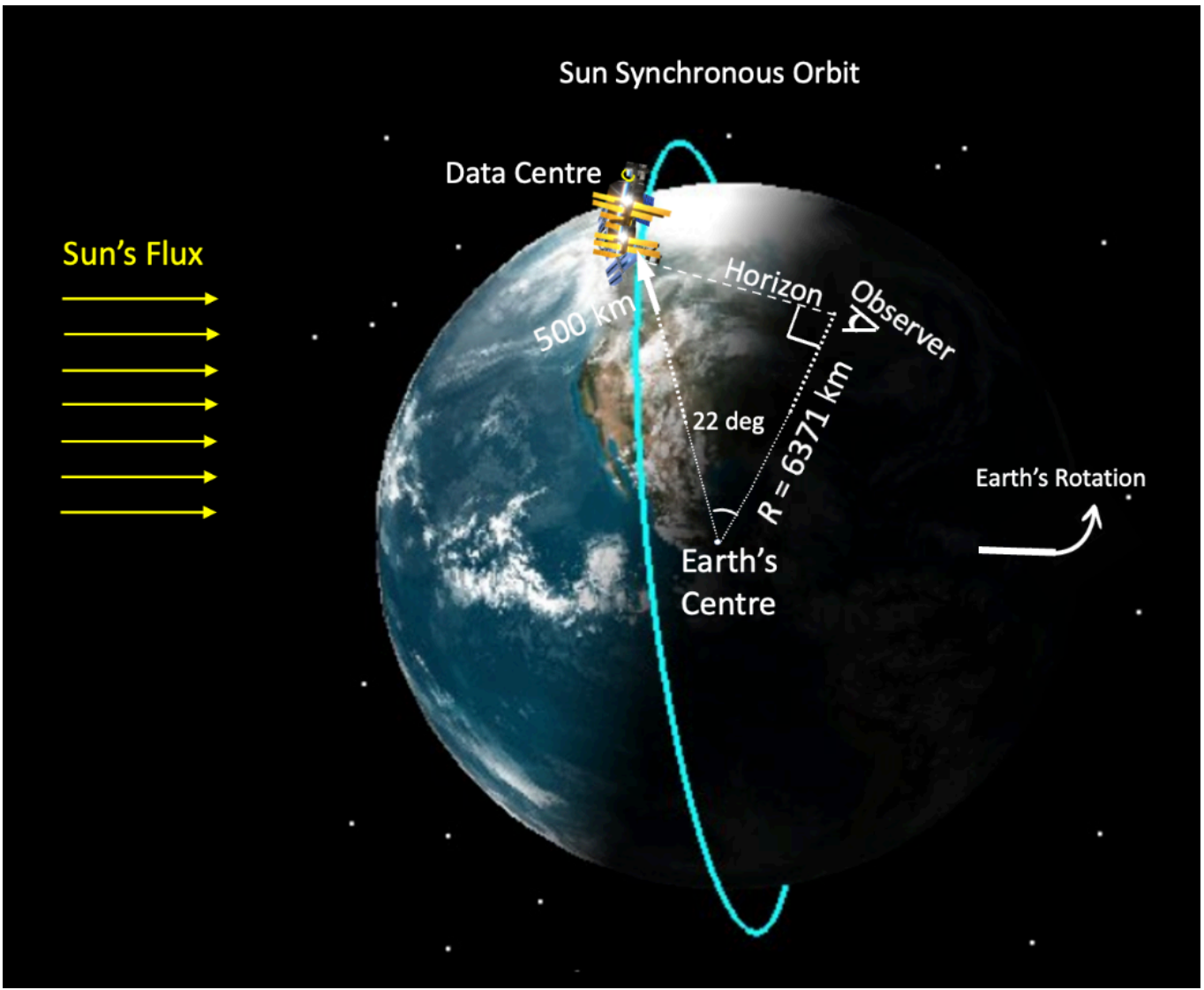

*Figure 1. The geometry showing the Sun-synchronous data centers visible at the horizon 90 minutes after sunset. At sunset, the data centers will reside in a swath passing directly overhead, from north-to-south (not shown). After 90 minutes, the observer is carried eastward causing the data centers, at 500 km altitude, to set on the observer's horizon (the dashed line). The 22 deg angle at the Earth's center follows from the right triangle. Data centers will inhabit the visible sky during 90 minutes both before and after sunset, and also 90 minutes before and after sunrise, every day. At high latitudes, the data centers will be visible all the time.*

Unlike other LEO satellites that vanish when they pass into the Earth's shadow, the sun-synchronous data centers will remain bathed in sunlight all the time. They only vanish when an observer is carried eastward by the rotation of the Earth, causing the satellites to set beneath the horizon. Symmetrically, the swath of data centers will rise in the east 90 minutes before sunrise and loom overhead right at sunrise.

3   SATELLITE BRIGHTNESS AND BLOCKAGE OF THE SKY

The kilometer-sized data centers in Sun-synchronous orbits will be above the horizon for all observers during at least ~6 hours every day, depending on latitude, including 3 hours surrounding sunset and 3 hours surrounding sunrise. During those 6 hours, their brightness and sky blockage will be substantial because their cross-sectional areas are 140 000 greater than current LEO satellites.

The large size of a 15 km² data centre brings several consequences, best estimated by scaling from current LEO satellites, such as the V2 Mini Starlink satellites, that have solar panels with a total area of ~105 m² (Mallama & Cole 2025; Mallama et al. 2025). The solar panels have low albedos of 0.05–0.20, and they reflect sunlight, appearing at visual magnitudes $g$ = 6 to 7.5 mag, just fainter than naked-eye visibility (Mallama 2023; Mallama & Cole 2025). When the current LEO satellites pass within ~1500 km of an observer (ground distance) during the 1.5 twilight hours after sunset or before sunrise, they appear above the horizon and interfere with astronomical observations made with telescopes both on the ground and space (Mróz et al. 2022; Kruk et al. 2023). During the middle ~6 hours of the night, surrounding midnight, most current LEO satellites reside in the Earth's shadow and are dark at visible, UV, and near-IR wavelengths. Even so, they still produce substantial pollution at mid-IR, far-IR, and radio wavelengths. Increased use of satellite laser communications at 1064 and 1550 nm, and glints from them, will further increase near-IR pollution.

It is rational to predict the brightness of the proposed 15 km² data centres by their cross-sectional areas relative to current ~105 m² of LEO satellites. This scaling implies the data centres will be ~140 000 times brighter, i.e. ~13 magnitudes brighter, implying that the data centres will have a brightness at magnitude g = -7 to -5.5 mag. Thus, they will be several times brighter than Venus (g ~ -4 at its brightest), and roughly 1/20 as bright as a quarter Moon (g ~ -10.5).

Certainly, as with LEO satellites, the solar panels of the proposed data centres will be oriented toward the Sun. However, as with LEO satellites, each structure will contain multiple surfaces that are not oriented perfectly toward the Sun, such as support beams and struts, rounded joints and corners, and tilted structural components all of which will scatter sunlight in all directions. Also, the spacecraft bus that contains the computing facility will reflect sunlight in all directions. Careful mitigating measures, such as visors, may reduce the reflected sunlight toward Earth. Such measures are already implemented in Starlink satellites designed to make them as faint as possible. If the new data centres match that level of mitigation, then the scaling computed above is appropriate. If however such mitigation is not implemented for the kilometer-sized structures, then they will be brighter than computed above.

The data centers will be visible in the sky during every late afternoon, within 90 minutes of sunset, and every morning for 90 minutes at sunrise. During the late afternoon, the sun-synchronous data centers will rise in the east reaching overhead at sunset. The fully illuminated solar panels will be visible to the naked eye against the blue sky.

More quantitatively, the albedos of solar panels are typically 10% to 15%. In comparison, the average albedo of the Moon is ~0.08 in maria and ~0.17 in the highlands, giving a weighted average lunar albedo of 0.11 (averaged over visible wavelengths). Thus, with similar albedos and both being extended objects, the surface brightness of the solar panels of the orbiting data centers would be nearly the same as that of the Moon. However, the orientation of the panels with normal vectors pointed toward the Sun and away from the ground, will cause the panels of the data centres to have lower surface brightness, as determined above. Still, at magnitude, g ~ -6, and 0.4 deg across (see below), with distinct specular reflection from specific components, the data-centre solar panels will be visible as faint, extended objects with bright spots before sunset and after sunrise for 1 ½ hrs.

The angular size of the solar panels can be estimated easily. Consider a representative orbital data center with solar-array linear size $D = 4$ km at altitude $h = 550$ km. The angular diameter $\delta$ (in radians) is given by, for small-angles,

$$\delta \approx \frac{D}{h} = \frac{4 \times 10^3}{5.5 \times 10^5} \approx 7.27 \times 10^{-3} \text{ rad} = 0.42°$$

This angular size of 0.42° is comparable to the angular diameter of the full Moon (~0.53°).

During the late afternoon and early morning hours, the data centers would be angularly nearly as large as the Moon, and roughly 1/20 as bright, making them easily visible in the blue sky. They would be odd, mechanical shapes, due to the complex solar panels and industrial computing centres associated with them. The resolved machines would be strewn across the blue sky in a swath from north to south, passing nearly overhead.

The data centres are so angularly large, 0.42° across, that they would temporarily block the view of stars, planets, and deep-sky objects for several seconds as they orbit the Earth every 90 minutes. Even to the naked eye, stars and planets would literally vanish. For the public, they would alter the visual character of the night sky. For example, when viewing the Orion constellation, a few of its 1$^{st}$ magnitude stars would disappear for a few seconds, despite no clouds. A crescent or quarter Moon would suddenly disappear as a data center moves in front. Saturn, Jupiter, Vega, and Sirius would commonly vanish. For professional transient surveys, such as by ZTF, Pan-STARRS, ARGUS Array, GOTO, ATLAS, and Rubin/LSST, the data centres will cause deep-sky objects to disappear and appear on time scales of seconds, but only happening during 1 ½ after sunset and before sunrise and near the horizon.

## 4  COLLISIONS

Collisions between satellites are inevitable, especially because many military satellites are launched and operated in secret. Discovering new satellites, determining their orbits, and detecting unannounced maneuvers remain major challenges, so many satellites are still uncatalogued.

A collision in LEO can produce thousands of debris fragments, which then strike other satellites and generate still more debris, triggering the Kessler syndrome (Kessler et al. 2010). In a runaway cascade, millions of fragments would fill the region 500–600 km above Earth, with no practical way to remove them. There is no effective means to clean such a vast orbital volume. Over decades, some debris would gradually lose altitude through atmospheric drag and eventually fall to Earth at random locations.

More than 1 million debris objects, 1–10 cm in size, already occupy low-Earth orbit (ESA 2025). With masses of roughly 1–1000 g and velocities near 7 km s$^{-1}$, they can seriously damage solar panels and data centres. Sun-synchronous satellites cross both poles, so their velocity vectors are nearly perpendicular to debris in equatorial orbits, producing relative speeds of about 10 km s$^{-1}$. A 1 kg impactor at that speed would deliver about $5 \times 10^7$ J, far exceeding the roughly 3,000 J of a 12-gauge shotgun or 7 mm rifle. Debris larger than a few centimetres would therefore severely damage solar panels and computing modules, while maneuvering kilometre-scale data centres to avoid such impacts would be difficult.

The International Space Station provides a useful example. NASA has recorded more than 1,400 impacts on the ISS, typically leaving pits, craters, or surface damage rather than causing major failures. Solar panels on kilometre-scale data centres would likewise need to remain functional despite centimetre-scale, million-joule punctures.

## 5  CONCLUSIONS

The proposed individual orbiting data centers, designed to deliver gigawatt-scale computational capacity for artificial intelligence, would require large solar arrays spanning several kilometers to harvest near-continuous solar energy. To be bathed in sunlight continuously, the computational facilities will reside in Sun-synchronous orbits at ~550 km altitude (LEO). Such architectures have already been proposed by a half dozen private entities, including FCC filings for approval, and have been studied by various governments. The motivation is to support 10- to 100-fold greater computing capability.

In twilight conditions (after sunset and before sunrise), the structures would be 100 times brighter than the brightest stars, with magnitudes, $g = -5.5$ to $-7.0$ mag. They would subtend angular diameters of ~0.4° nearly as large as the Moon. Their industrial geometries would be resolved and discerned by the naked eye, and would cause them to occult stars and planets for durations of several seconds during transit. In daylight hours near dawn and dusk, the orbiting computational facilities would appear as a series of industrial objects visibly moving across the blue sky from north to south.

Prior to widespread implementation, the astronomical and societal consequences of such large machines in LEO will require rigorous, multidisciplinary assessment. The visible encroachment into the daytime and nighttime sky would affect everyone, including communities, countries, and cultures having no direct role in the computers.

Across history, the natural wonder of the sky has served as a common reference point for diverse societies, and likely even for early hominins. Large, bright artificial structures in orbit would significantly alter this shared natural environment. For astronomers, policymakers, and communities worldwide, such orbiting computing facilities would obstruct and alter the natural sky seen by everyone on Earth, modifying a natural source of thoughtfulness about humanity's role in the cosmos.

**ACKNOWLEDGEMENTS**

This work benefitted from valuable communications with Jacob Bouchard, John Gertz, Hasan Bahcivan, Maryanne Lynch, Kevin Krisciunas, and Susan Kegley. We thank the team at Space Laser Awareness for outstanding support.

**DATA AVAILABILITY**

This paper is based on no new data.

This paper was typeset from Microsoft WORD document prepared by the author.